\documentclass[aps,prl,tightenlines,twocolumn,superscriptaddress]{revtex4-1}

\usepackage{amsmath,bm,mathrsfs,amssymb,extarrows}
\usepackage{graphicx,subfigure,float}
\usepackage{multirow, booktabs}
\usepackage{xcolor,enumerate}
\usepackage{paralist}
\usepackage[colorlinks, linkcolor=red, anchorcolor=blue, citecolor=purple]{hyperref}

\newcommand{\beq}{\begin{equation}}
\newcommand{\eeq}{\end{equation}}
\newcommand{\beqs}{\begin{equation*}}
\newcommand{\eeqs}{\end{equation*}}
\newcommand{\ii}{\mathrm{i}}

\def\Rb87{^{87}\rm{Rb}}
\def\Na23{^{23}\rm{Na}}
\def\Cr52{^{52}\rm{Cr}}
\makeatletter

\newcommand{\Rmnum}[1]{\expandafter\@slowromancap\romannumeral #1@}
\makeatother

\begin{document}
\title{Quantum Phases of Time Order in Many-Body Ground States}
\author{Tie-Cheng Guo}
\email{gtc16@mails.tsinghua.edu.cn}
\affiliation{State Key Laboratory of Low Dimensional Quantum Physics,\\ Department of Physics, Tsinghua University, Beijing 100084, China}
\author{Li You}
\email{lyou@mail.tsinghua.edu.cn}
\affiliation{State Key Laboratory of Low Dimensional Quantum Physics,\\ Department of Physics, Tsinghua University, Beijing 100084, China}
\affiliation{
Frontier Science Center for Quantum Information, Beijing, China}

\date{\today}

\begin{abstract}
Understanding phases of matter is of both fundamental and practical importance.
Prior to the widespread appreciation and acceptance of topological order,
the paradigm of spontaneous symmetry breaking,
formulated along the Landau-Ginzburg-Wilson (LGW) dogma, is central to understanding phases
 associated with order parameters of distinct symmetries and transitions between phases.
This work proposes to identify ground state phases of quantum many-body system
in terms of \textit{time order}, which is operationally defined by
the appearance of nontrivial temporal structure in the
two-time auto-correlation function of a symmetry operator (order parameter).
As a special case, the (symmetry protected) {\it time crystalline order} phase detects
continuous time crystal (CTC).
Time order phase diagrams for spin-1 atomic Bose-Einstein condensate (BEC)
and quantum Rabi model are fully worked out.
Besides time crystalline order, the intriguing phase of time functional order 
is discussed in two non-Hermitian interacting spin models.
\end{abstract}
\maketitle

A consistent theme for studying many-body system, particularly in condensed matter physics,
concerns the classification of phases and their associated phase transitions
\cite{wenbook2004qf, Fradkin2013field, Sachdev1999qpt}.
In the celebrated Landau-Ginzburg-Wilson (LGW) paradigm \cite{lgw1999, wilson1974renormalization},
spontaneous symmetry breaking plays a central role with order parameters characterizing
different phases of matter possessing respective broken symmetries.
Other schemes for classifying phases as well as their associated transitions are, 
however, beyond the Landau-Ginzburg-Wilson paradigm,
which are by now well accepted since first established
 decades ago \cite{senthil2004deconfined, wen1989vacuum, wen1990toporigid}.
For example, topological order, which classifies gapped quantum many-body system
constitutes a topical research direction \cite{wen1989vacuum, wen1990toporigid, wen2017rmpcollo, wen2019choreographed}.
Our current understanding categories gapped systems into gapped liquid phases \cite{zeng2015gapped} and
gapped non-liquid phases,
with the former broadly including phases of topological order \cite{wen1989vacuum, wen1990toporigid},
symmetry enriched topological order \cite{wen2002quantumorder, chen2015anomalous, cheng2017exactly, heinrich2016symmetry}, and
symmetry protected trivial order \cite{gu2009tensor, chen2011two, chen2013symmetry},
while the recently discussed fracton phases \cite{wilbur2019univer, shirley2018fracton, vijay2016fracton}
belongs to the latter of
gapped non-liquid phases.

Temporal properties of phases are also worthy of investigations
as exemplified by many recent studies \cite{Sacha2017temporaldisorder, sacha2015anderson, Wilczek2012qtc}.
For instance, time crystal (TC) or perpetual temporal dependence
in a many-body ground state that breaks spontaneously time translation symmetry (TTS),
constitutes an exciting new phenomenon.
First proposed by Wilczek \cite{Wilczek2012qtc}
for quantum systems
and followed by Shapere and Wilczek \cite{Wilczek2012ctc}
for classical systems in 2012,
TC in their original sense is unfortunately ruled out by Bruno's no-go theorem the following year \cite{Bruno2013impossibility, Nozi_res_2013}.
Watanabe and Oshikawa (WO) reformulate the idea of quantum TC \cite{WO2015absence},
and present a refined no-go theorem for many-body systems without too long-range interactions \cite{WO2015absence}.
Most recent efforts on this topic are directed towards non-equilibrium discrete/Floquet TC breaking
discrete TTS \cite{Sacha2015modeling, von2016absolutesta, else2016floquettc, Khemani2016phasestructure, NYYao2017rigidity, Norman2020reviewdtc},
particularly in systems with disorder that facilitate many-body localizations \cite{von2016absolutesta, NYYao2017rigidity},
in addition to clean systems \cite{Russomanno2017LMGclean, Huang2018cleancold, Fan2020Rydberg, machado2020longrange}.
Ongoing studies are further extended to open systems with Floquet driving in the presence of dissipation \cite{Gong2018cavitycircuit, Gambetta2019absencesymmetry, riera2019time, Lazarides2020dissipativedtc, cosme2019shakenca}, with experimental investigations reported for a variety of systems \cite{2017trappedions, 2017diamonddipolor, rovny2018nmr, rovny2018orderdi, pal2018tem, autti2018ob, smits2018obser}.
A recent study addresses TC and its associated physics along imaginary time axis \cite{caizi2020iTC}.

We introduce \textit{time order} in this work, as the essential element for a new perspective
to identify and categorize quantum many-body phases,
based on different ground state temporal patterns.
Each quantum many-body Hamiltonian $\hat{H}$ comes with its evolution
or time translation operator $e^{-i\hat{H}t}$.
When continuous time translation symmetry is broken for operator $e^{-i\hat{H}t}$,
akin to the breaking of continuous spatial translation symmetry
for operator $e^{-i\vec k\cdot \vec r}$,
time crystals arise in direct analogy to spatial crystals \cite{Wilczek2012qtc}.
The message we hope to convey here in this study is rooted on the dual between $\hat{H}$ 
and $e^{-i\hat{H}t}$, which we argue quite generally establishes a solid foundation
for time order and provides further information concerning
ground state quantum phases based on time domain properties.
Different quantum many-body states with the same temporal patterns
are classified into the same time order phases, of which
continuous TC (CTC), a ground state with periodic time dependence breaking continuous TTS
as originally proposed in Refs. \cite{Wilczek2012qtc, Wilczek2012ctc},
belongs to one of them.

We will adopt the WO definition of CTC based on two-time auto-correlation function of an operator.
First outlined in the now famous no-go theorem work \cite{WO2015absence}, 
it establishes a general and rigorous subtype of CTC: the WO CTC.
Recently, Kozin and Kyriienko claim to have realized such a genuine ground state CTC
in a multi-spin model with long-range interaction \cite{Kozin2019qtcLongRange},
buttressing much confidence to the search for exotic CTCs.
The operational definition for time order we introduced encompasses WO CTC as one type of time order phases.
We will also explore and elaborate a variety of possible exotic phases.

\section{Results}
\subsection{Time order}
We argue that ground state temporal properties of 
a quantum many-body system can be used to characterize or classify its phases.
Hence, the concept of \textit{time order} can be introduced analogous to
an order parameter by bestowing it in the non-trivial temporal dependence.
To exemplify the essence of the associated physics,
we shall present an operational definition for {\it time order} and
accordingly work out the exhaustive list of all allowed phases.
According to the WO proposal \cite{WO2015absence},
a witness to CTC is the following two-time (or unequal time)
auto-correlation function (with respect to ground state)
\beq \lim_{V\rightarrow\infty} \langle\hat{\Phi}(t)\hat{\Phi}(0)\rangle/V^2 \equiv f(t),
\eeq
for operator $\hat{\Phi}(t)\equiv\int_Vd^Dx\hat{\phi}(\vec{x}, t)$ defined as an integrated order parameter (over $D$-spatial-dimension), or analogously the volume averaged one,
\beq f(t)=\lim_{V\rightarrow\infty} \langle\hat{\phi}(t)\hat{\phi}(0)\rangle,\eeq
with $\hat{\phi}(\vec{x}, t)$ the corresponding local order parameter density 
operator $\hat{\phi}\equiv\hat{\Phi}/V$.

If $f(t)$ is time periodic,
the system is in a state of CTC. This can be reformulated
into an explicit operational protocol by introducing twisted vector.
 \textit{For a quantum many-body system with energy eigen-state $|\psi_i\rangle$,
if there exists a coarse-grained Hermitian order parameter $\hat{\phi}$, $\hat{\phi}|\psi_i\rangle$ is called the eigen-state twisted vector;
More generally, if $\hat{\phi}$ is non-Hermitian, $\hat{\phi}|\psi_i\rangle$
(or $\hat{\phi}^\dagger|\psi_i\rangle$) will be called the
right (or left) eigen-state twisted vector.}

The orthonormal set of eigen-wavefunctions
$|\psi_i\rangle~(i=0,1,2,\cdots)$ for 
a system described by Hamiltonian $\hat{H}$ is arranged 
in increasing eigen-energies $\epsilon_i$ with $i=0$ denoting the ground state.
When the coarse-grained order parameter $\hat{\phi}$ is Hermitian, the ground state twisted vector $|v\rangle$ can be expanded $|v\rangle\equiv\hat{\phi}(0)|\psi_0\rangle=\sum_{i=0}^\infty a_i |\psi_i\rangle$ into the eigen-basis.
With the help of the Schr\"odinger equation $i{\partial|\psi(t)\rangle}/{\partial t} = \hat{H} |\psi(t)\rangle$
($\hbar=1$ assumed throughout) for the system wave function $|\psi(t)\rangle$, we obtain
\begin{eqnarray}\label{hermitianf}
	f(t) 
	&=& \lim_{V\rightarrow\infty}\langle\psi_0|e^{i\hat{H}t}\hat{\phi}(0)e^{-i\hat{H}t}\hat{\phi}(0)|\psi_0\rangle\nonumber\\
	&=& \lim_{V\rightarrow\infty}e^{i\epsilon_0 t}\langle v| e^{-i\hat{H}t} |v\rangle\nonumber\\
	&=& \lim_{V\rightarrow\infty}\sum_{j=0}^\infty \eta_j e^{-i(\epsilon_j-\epsilon_0)t},
\end{eqnarray}
where $\eta_j\equiv |a_j|^2$ denote weights of the ground state twisted vector,
$\eta_0$ the corresponding ground state weight, and $\eta_j$ (with $j>0$) the excited state weight.

When the coarse-grained order parameter $\hat{\phi}$ is non-Hermitian, we use $|v^{(l)}\rangle$ and $|v^{(r)}\rangle$ to
denote respectively the left and right ground state twisted vectors and expand them analogously
in the eigen-basis
to arrive at $|v^{(l)}\rangle\equiv\hat{\phi}^\dagger|\psi_0\rangle = \sum_{i=0}^\infty b_i|\psi_i\rangle$ and $|v^{(r)}\rangle\equiv\hat{\phi}|\psi_0\rangle=\sum_{i=0}^\infty a_i|\psi_i\rangle$. In this case, we find
\begin{eqnarray}
\label{nonhermitianf}
	f(t) 
	&=& \lim_{V\rightarrow\infty}\langle\psi_0|e^{i\hat{H}t}\hat{\phi}(0)e^{-i\hat{H}t}\hat{\phi}(0)|\psi_0\rangle\nonumber\\
	&=& \lim_{V\rightarrow\infty}e^{i\epsilon_0 t}\langle v^{(l)}| e^{-i\hat{H}t} |v^{(r)}\rangle\nonumber\\
	&=& \lim_{V\rightarrow\infty}\sum_{j=0}^\infty \eta_j e^{-i(\epsilon_j-\epsilon_0)t},
\end{eqnarray}
with $\eta_j\equiv b_j^*a_j$ weights of the ground state twisted vector instead.
Similarly $\eta_0$ and $\eta_j$ ($j>0$)
denote respectively ground and excited state weights.

Given an order parameter $\hat{\phi}$, quite generally
$f(t)$ is a sum of many harmonic functions with amplitudes $\eta_j$ and characteristic frequencies $\omega_j\equiv\epsilon_j-\epsilon_0$.
Nontrivial time dependence of the two-time auto-correlation function is thus imbedded in
the energy spectra of $H$ as well as in the weights of the ground state twisted vector.
For CTC order to exist, one of the excited state weights must be non-vanishing,
or in rare cases, $f(t)$ can include harmonic terms of commensurate frequencies.

If $f(t)$ is a constant, the time dependence will be trivial.
However, a subtlety appears when $f(t)$ is vanishingly small with respect to system size.
Since what we are after is the system's explicit temporal behavior or time dependence,
which is easily washed out to $f(t)=0$ by a vanishing norm of the twisted vector.
Such a difficulty can be mitigated by multiplying system volume $V$,
\textit{i.e.}, using the twisted vector $|v\rangle\rightarrow V|v\rangle$
to check if the correlation for the bulk order parameter $F(t)\equiv V^2f(t)$ exhibits temporal dependence, or vanishes.
\beq F(t)=\lim_{V\rightarrow\infty} \langle\hat{\Phi}(t)\hat{\Phi}(0)\rangle.\eeq
When $f(t)=0$ but $F(t)$ remains a periodic function, the system can still be considered a CTC.
Such a remedy surprisingly captures the essence of generalized CTC of Ref.\cite{dieter2019heisenberg}.

\begin{table*}
	\centering
	\caption{Classification of the ground state phases for a quantum many-body system}
	\label{table:torder2}
	\begin{ruledtabular}
		\begin{tabular}{c|l|l}
			\multicolumn{2}{c|}{Phase}  & Property of two-time auto-correlator\\
			\colrule
			\multicolumn{2}{l|}{Time trivial order} & $f(t)={\rm const.}\neq 0$ or $f(t)=0, F(t)={\rm const.}$\\ 
			\cline{1-3}
			\multirow{3}{*}{Time order} & Time crystalline order & $f(t)$ is periodic and nonvanishing  \\
			& Time quasi-crystalline order & $f(t)$ is quasiperiodic with beats from two incommensurate frequencies\\
			& Time functional order & $f(t)$ is aperiodic  \\
			& Generalized time crystalline order & $f(t)=0$, $F(t)$ is periodic and nonvanishing  \\
			& Generalized time quasi-crystalline order & $f(t)=0$, $F(t)$ contains beats from two incommensurate frequencies \\
			& Generalized time functional order & $f(t)=0$, $F(t)$ is aperiodic\\
		\end{tabular}
	\end{ruledtabular}
\end{table*}

The analysis presented above can be directly 
extended to excited states \cite{Syrwid2017ExcitedContinuous}.
It is also straightforwardly applicable to non-Hermitian systems, 
as long as a plausible ``ground state" can be identified,
for example, by requiring its eigen-energy to possess the largest imaginary part or the smallest norm.
Denoting the imaginary part of energy eigen-value $E_i$ as $\text{Im}(E_i)$,
a prefactor $\propto e^{\text{Im}(E_i)t}$ then arises in the auto-correlation function,
leading to unusual time functional order in the classification of time order.

Therefore, quantum many-body phases can be classified
according to {\it time order}. The
two-time auto-correlation function based complete operational procedure for classifying time order thus extends the definition of WO CTC in Ref. \cite{WO2015absence}.
Our central results can be simply stated in the following:
\textit{
If $f(t)$ exhibits nontrivial time dependence, time order exists.
If $f(t)=0$, but $F(t)$ displays nontrivial time dependence instead,
generalized time order exists.}

More specifically, if $f(t) = {\rm const.}$ is nonzero,
the system exhibits time trivial order.
The same applies when $f(t)=0$
and $F(t)={\rm const.}$.
For all other situations, nontrivial time order prevails.
A complete classification
for all time order ground state phases is shown in Table \ref{table:torder2},
according to the temporal behaviors
of their auto-correlation functions $f(t)$ or $F(t)$. As shown in the
Supplementary Material (SM), the above discussion and classification on time order can be extended to
finite temperature systems as well.

The operational procedure outlined above presents a straightforward
approach for detecting {\it time order},
albeit with reference to an order parameter operator.
Hence more appropriately, this approach should be called \textit{order parameter assisted time order}
or \textit{symmetry-based (or -protected) time order}, to emphasize its reference
to symmetry order parameter
of a quantum many-body system.
The twisted vector facilitates easy calculations to distinguish between
different time order phases from time trivial ones,
as we illustrate below in terms of a few concrete examples.
It is reasonable to expect that transitions between different time order phases can occur,
reminiscent of phase transitions in the LGW spontaneous symmetry breaking paradigm.

\subsection{Time order phase in a spin-1 atomic condensate}
A spin-1 atomic Bose-Einstein condensate (BEC)
under single spatial mode approximation (SMA) \cite{Law1998quantumspins, pu1999spinmixing, yi2002single}
is described by the following Hamiltonian
	\begin{eqnarray}
  \hat H &= &\frac{c_2}{2N}\left[\left(2\hat N_0 -1\right)\left(\hat N-\hat N_{0}\right)+2\left(\hat a_1^\dag\hat a_{-1}^\dag\hat a_0\hat a_0+\text{h.c.}\right)\right]\nonumber\\
&&   -p \left(\hat N_1 - \hat N_{-1}\right)
	+ \,q\,\left(\hat N_1 + \hat N_{-1}\right)\,,\label{eq:Hamil0}
\end{eqnarray}
where $\hat{a}_{m_F}(m_F = 0, \pm 1)$ ($\hat{a}_{m_F}^\dag$) denotes the annihilation (creation) operator
for atom in the ground state Zeeman manifold $|F = 1, m_F\rangle$
with corresponding number operator $\hat{N}_{m_F} = \hat{a}_{m_F}^\dagger \hat{a}_{m_F}$.
The total atom number $\hat{N} = \hat{N}_1 + \hat{N}_0 + \hat{N}_{-1}$ is conserved.
$p$ and $q$ are linear and quadratic Zeeman shifts that can be tuned independently \cite{luo2017deterministic},
while $c_2$ describes the strength of spin exchange interaction.

The validity of this model is well established based on extensive
theoretical \cite{chang2005coherent, zhang2010localization, guzman2011long, xue2018kz}
and experimental \cite{chang2004observation, anquez2016quanKZ, luo2017deterministic, qiu2020observation} studies of spinor BEC over the years.
The fractional population in spin states $|1, 1\rangle$ and $|1, -1\rangle$,
$\hat{n}_{\rm sum}\equiv N_{\rm sum}/{N}$, with $N_{\rm sum}={\hat{N}_1+\hat{N}_{-1}}=N-N_0$,
is often chosen as an order parameter \cite{damski2007dynamics,lamacraft2007quench, anquez2016quanKZ, xue2018kz} with $N$ assuming the role of system size.
The ground state twisted vector then becomes $|v\rangle\equiv\hat{n}_{\rm sum}|\psi_0\rangle$, and
\begin{eqnarray}
	f(t) &=&\lim_{N\rightarrow\infty}\langle\hat{n}_{\rm sum}(t)\hat{n}_{\rm sum}(0)\rangle,\\
	F(t) &=& \lim_{N\rightarrow\infty}
	\left\langle\hat{N}_{\rm sum}(t)\hat{N}_{\rm sum}(0)\right\rangle.
\end{eqnarray}

\begin{figure}[htbp]
	\centering
	\includegraphics[scale=1]{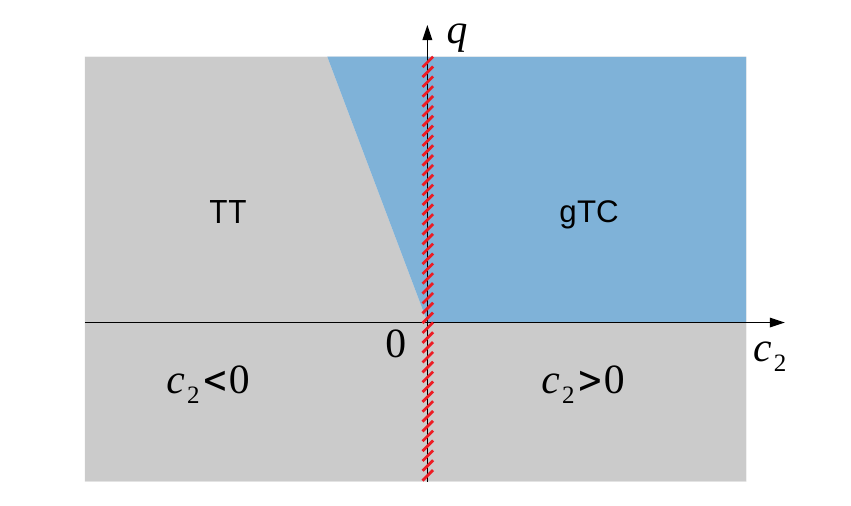}
	\caption{Time order phase diagram for spin-1 atomic BEC, 
		where TT and gTC respectively denote time trivial and generalized time crystalline order.
		The region of (hashed) line segments surrounding $c_2=0$ for noninteracting system is to be excluded.}
	\label{fig:phaseDiagram}
\end{figure}

We will concentrate on the zero magnetization $F_z=0$ subspace
and employ exact diagonalization (ED) to calculate eigen-states.
$p=0$ is assumed since $F_z$ is conserved.
Figure \ref{fig:phaseDiagram} illustrates
the system's complete time order phase diagram.
For ferromagnetic interaction $c_2<0$ as with $\Rb87$ atoms, the critical quadratic Zeeman shift $q/|c_2|=2$
splits the whole region into time trivial order (TT) phase for smaller $q$ that observes TTS,
and generalized time crystalline (gTC) order phase for $q/|c_2|>2$
where TTS is spontaneously broken.
The latter (gTC phase) is found to coincide with the polar phase \cite{xue2018kz}.
Limited by available computation resources,
the system sizes we explored with ED remain moderate which prevent us from
mapping out the finer details in the immediate neighborhood of $q=2|c_2|$.
Further elaboration of time order properties in this region is therefore needed.
On the other hand, for antiferromagnetic interaction $c_2>0$ with $\Na23$ atoms,
we find $q=0$ separates TT phase from gTC order.
We note here that $q=2|c_2|$ is the second-order quantum phase transition (QPT) critical point between the polar phase and the broken-axisymmetry phase of the ferromagnetic spin-1 BEC, while $q=0$ corresponds to the first-order QPT critical point for antiferromagnetic interaction .

More detailed discussions including the dependence of time order phases on system size,
possible approaches to detect them, and extension to thermal state phases
 can be found in the
SM.

\subsection{Time order phase diagram for quantum Rabi model}
As a second example, we consider time order phases
of the quantum Rabi model described by the Hamiltonian
\beq \hat{H}_\text{Rabi} =  \omega_0 \hat{a}^\dagger \hat{a} + \frac{\Omega}{2}\hat{\sigma}_z - \lambda(\hat{a}+\hat{a}^\dagger)\hat{\sigma}_x,\eeq
where $\hat{\sigma}_{x,z}$ is Pauli matrix of a two-level system
(transition frequency $\Omega$),
$\hat{a} (\hat{a}^\dagger)$ is the annihilation (creation) operator
for a single bosonic field mode (of frequency $\omega_0$),
and $\lambda$ is their coupling strength.

It is known that the above model exhibits a QPT to a superradiant state,
despite of its simplicity \cite{Hwang2015qptRabi}.
The transition
occurs at the critical point $g_c\equiv1$, with the
dimensionless parameter $g\equiv 2\lambda/\sqrt{\omega_0\Omega}$.
The equivalent thermodynamic limit is approached by taking
$\Omega/\omega_0\rightarrow\infty$.
According to the studies in Ref. \cite{Hwang2015qptRabi},
an almost exact effective low-energy Hamiltonian for the normal phase ($g<1$) is given by
\beq \hat{H}_\text{np} = \omega_0 \hat{a}^\dagger \hat{a} - \frac{\omega_0 g^2}{4} (\hat{a}+\hat{a}^\dagger)^2 - \frac{\Omega}{2},\eeq
whose low-energy eigen-states are
$|\phi_\text{np}^m(g)\rangle = \hat{\mathcal{S}}[r_\text{np}(g)]|m\rangle|\downarrow\rangle$
for $g\leq1$, with
$\hat{\mathcal{S}}[x]=\exp[{x}(\hat{a}^{\dagger2}-\hat{a}^2)/{2}]$ and $r_\text{np}(g)=-[\ln(1-g^2)]/4$,
and the energy eigen-values are
$E_\text{np}^m(g) = m\epsilon_\text{np}(g) + E_\text{G,np}(g)$,
with $\epsilon_\text{np}(g)=\omega_0\sqrt{1-g^2}$ and $E_\text{G,np}(g)=[{\epsilon_\text{np}(g)-\omega_0}]/{2}-{\Omega}/{2}$.
For the supperadiant phase ($g>1$), the effective low energy Hamiltonian becomes
\beq \hat{H}_\text{sp} = \omega_0 \hat{a}^\dagger \hat{a} - \frac{\omega_0}{4g^4}(\hat{a}+\hat{a}^\dagger)^2
-\frac{\Omega}{4}(g^2+g^{-2}),\eeq
whose eigen-states are given by
$|\phi_\text{sp}^m(g)\rangle_{\pm} = \hat{\mathcal{D}}[\pm\alpha_g]\hat{\mathcal{S}}[r_\text{sp}(g)]|m\rangle|\!\downarrow^\pm\rangle$,
with $r_\text{sp}(g) = -[\ln(1-g^{-4})]/4$, $\alpha_g = \sqrt{(\Omega/4g^2\omega_0)(g^4-1)}$, and $\hat{\mathcal{D}}[\alpha] = e^{\alpha(\hat{a}^\dagger - \hat{a})}$.
The displacement-dependent spin states are
$|\!\downarrow^\pm\rangle = \mp \sqrt{(1-g^{-2})/2}|\!\uparrow\rangle
+ \sqrt{(1+g^{-2})/2}|\!\downarrow\rangle$, while the energy eigen-values take the form
$E_\text{sp}^m(g) = m\epsilon_\text{sp}(g) + E_\text{G,sp}(g)$,
with $\epsilon_\text{sp}(g)=\omega_0\sqrt{1-g^{-4}}$ and $E_\text{G,sp}(g)=[{\epsilon_\text{sp}(g)-\omega_0}]/{2}-{\Omega}(g^2+g^{-2})/{4}$.
More details can be found in the SM of Ref. \cite{Hwang2015qptRabi}.

For this model,
the scaled average cavity photon number $\hat{n}_c={\omega_0}\hat{a}^\dagger \hat{a}/{\Omega}$
is a suitable order parameter with $\Omega/\omega_0$ assuming the role of system size.
The corresponding bulk order parameter then becomes $\hat{N}_c=\hat{a}^\dagger \hat{a}$
or the average cavity photon number, and
\beq\begin{split}
	f(t) & = \lim_{\Omega/\omega_0\rightarrow\infty}
	\left\langle \hat{n}_c(t)\hat{n}_c(0)\right\rangle, \\
	F(t) & = \lim_{\Omega/\omega_0\rightarrow\infty}
	\left\langle \hat{N}_c(t)\hat{N}_c(0)\right\rangle. \\
\end{split}\eeq

For $g<1$, we find
\beq\begin{split}
	f(t) & = 0, \\
	F(t) & = \eta_0 + \eta_2 e^{-i(2\epsilon_{\rm np})t},
\end{split}\eeq
respectively, where $\eta_0=\sinh^4(r_{\rm np})$ and $\eta_2 = \cosh^2(r_{\rm np})\sinh^2(r_{\rm np})$.
For $g>1$, we obtain
\beq\begin{split}
	f(t) & = \frac{(g^2-g^{-2})^2}{16}.
\end{split}\eeq

\begin{figure}[htbp]
\centering
\includegraphics[scale=1]{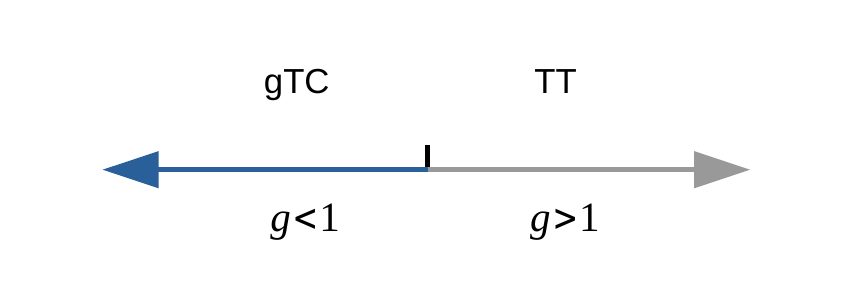}
\caption{Time order phase diagram for the quantum Rabi model, where
	TT and gTC respectively denote time trivial and generalized time crystalline order.}
\label{fig:Rabi}
\end{figure}

The time order phase diagram is shown in Fig. \ref{fig:Rabi}.
When $g<1$, the system ground state corresponds to a generalized time crystalline order phase,
 while the system exhibits time trivial order when $g>1$.
Despite of such a simple model composed of a two-level system and a bosonic field mode,
the ground state of the quantum Rabi model displays intriguing temporal phase structure
 accompanied by a finite-component quantum phase transition.

\subsection{Non-Hermitian many-body interaction model}
Finally we consider two effective models
with many-body spin-spin interaction and non-Hermitian effects.
The first is described by Hamiltonian
\begin{eqnarray}
\hat{H} &=& -\frac{1}{N(N-1)}\nonumber\\
&&\sum_{1\leq i<j\leq N}(\lambda+\ii\gamma)\sigma_1^x\sigma_2^x
\cdots\sigma_{(i)}^y\cdots\sigma_{(j)}^y\cdots\sigma_N^x, \hskip 24pt
\label{eq:nh1}
\end{eqnarray}
with two $\sigma^y$ operators at sites $i$ and $j$ in a string of
otherwise $\sigma^x$ $N$-body spin interaction. ${1}/[{N(N-1)}]$ is the normalization factor,
$\lambda$ is the spin interaction strength, and $\gamma$ represents an effective dissipation rate. $\lambda>0$ and $\gamma\geq0$ are both real numbers.

We observe that the Greenbergenr-Horne-Zeilinger (GHZ) states
\beq |G_\pm\rangle = \frac{1}{\sqrt{2}}(|0\rangle^{\otimes N}\pm|1\rangle^{\otimes N})\eeq
correspond to two nondegenerate system eigen-states with eigen-energies $\pm(\lambda+\ii\gamma)/2$.
The spectra of this model system is bounded inside the circle of radius ${\sqrt{\lambda^2+\gamma^2}}/{2}$ in the complex plane.
The eigen-state whose eigen-value has the largest imaginary part ia taken as the ground state, 
or $|{\rm GS}\rangle = |G_+\rangle$ with eigen-energy $\epsilon_0 = (\lambda+ \ii\gamma)/2$.
The highest excited state is $|G_-\rangle$, whose corresponding eigen-energy is $\epsilon_{2^N-1} = -(\lambda+\ii\gamma)/2$.

An appropriate order parameter operator in this case becomes the average magnetization $\hat{m} = \sum_{i=1}^N \sigma_i^z/N$. The twisted vector becomes $|v\rangle = \hat{m}|{\rm GS}\rangle = |G_-\rangle$,
and the auto-correlator can be easily worked out to be
$f(t) = \lim_{N\rightarrow\infty}\langle\hat{m}(t)\hat{m}(0)\rangle = e^{-i\lambda t} e^{\gamma t}$.
When $\gamma=0$, the system ground state exists time-crystalline order phase and corresponds to a continuous time crystal \cite{Kozin2019qtcLongRange}. When $\gamma\not=0$, the system exhibits time functional order,
with an exploding $f(t)$ as time evolves.

A second non-Hermitian model Hamiltonian is given by
\begin{eqnarray}
\hat{H} &=&
(\lambda+\ii\gamma) \left(\sigma_1^x\sigma_2^x\cdots\sigma_{[N/2]}^x
-\sigma_{[N/2]+1}^x\cdots\sigma_N^x\right)\nonumber\\
&&
-\sum_{j=1}^N\sigma_j^z\sigma_{j+1}^z,
\label{nonH}
\end{eqnarray}
where $[\cdot]$ denotes the integer part, $\bm\sigma_{N+1}\equiv\bm\sigma_1$ corresponds to the periodic boundary condition,
$\lambda$ and $\gamma$ are spin-string interaction strength
and dissipation strength respectively as in the previous model, both of them are real.
This Hamiltonian contains $[(N+1)/2]$-body interaction terms
and supports GHZ state $|G_+\rangle$ as a non-degenerate excited state \cite{Facchi2011GHZ}
with eigen-energy $\epsilon_+ = -N$.
The other two eigen-states of concern are
$|\Psi^{(\pm)}\rangle\equiv ({\alpha_1}|G_-\rangle + {\alpha_2}|\tilde{G}_{-,\mathcal{I}}\rangle)/{\sqrt{|\alpha_1|^2+|\alpha_2|^2}}$
with $\alpha_1=1$ and $\alpha_2 = {-(N+\epsilon^{(\pm)})}/{2(\lambda+\ii\gamma)}$, where
\beq\begin{split}
|\tilde{G}_{-,\mathcal{I}}\rangle
=&\frac{1}{\sqrt{2}}\large(
|0\rangle_1\cdots|0\rangle_{[N/2]}|1\rangle_{[N/2]+1}\cdots|1\rangle_N\\
&-|1\rangle_1\cdots|1\rangle_{[N/2]}|0\rangle_{[N/2]+1}\cdots|0\rangle_N
\large).
\end{split}\eeq
The eigen-energies for $|\Psi^{(\pm)}\rangle$ are given by
$\epsilon^{(\pm)}=-N+2\pm2\sqrt{1+(\lambda+ \ii\gamma)^2}$,
with
more details of the derivation given in SM.
For the same order parameter operator $\hat{m}$,
we find $\hat{m}|\Psi_0\rangle \overset{N\rightarrow\infty}{\longrightarrow} {\alpha_1}
|G_+\rangle/{\sqrt{|\alpha_1|^2+|\alpha_2|^2}}$.

At $\gamma=0$, the above non-Hermitian Hamiltonian (\ref{nonH}) reduces to a Hermitian one,
 whose ground state $|\Psi_0\rangle$ corresponds to the one with smaller $\epsilon$
from $|\Psi^{(-)}\rangle$ and $|\Psi^{(+)}\rangle$, or $\epsilon_0=-N-2(\sqrt{1+\lambda^2}-1)$.
The ground state $|\Psi_0\rangle$ for this non-Hermitian system
is therefore chosen from $|\Psi^{(-)}\rangle$ or $|\Psi^{(+)}\rangle$
to be the one that deforms into the right Hermitian case one when $\gamma$ approaches zero.
However, the criteria for the ground state energy $\epsilon_0$ corresponds to choosing the smaller one from $\epsilon^{(\pm)}$ when $\epsilon$ is real and choosing the one with the larger imaginary part when $\epsilon$ is complex.

Therefore we directly obtain
\beq f(t) = \lim_{N\rightarrow\infty} \langle \hat{m}(t)\hat{m}(0)\rangle
 = \frac{|\alpha_1|^2}{|\alpha_1|^2+|\alpha_2|^2}
 e^{-\ii(\epsilon_+-\epsilon_0)t}.
\eeq
When $\lambda\not=0$ and $\gamma\not=0$, the system exists in
 \textit{time functional order phase}, again results from the non-Hermitian Hamiltonian.
When $\lambda\not=0$ but $\gamma=0$, the auto-correlation function reduces to
\beq f(t) = \frac{1}{2}(1+\frac{1}{\sqrt{1+\lambda^2}})
e^{-2\ii(\sqrt{1+\lambda^2}-1)t},\eeq
as for a genuine time crystal of the WO type
exhibiting time crystalline order.
When $\lambda=0$ and $0<|\gamma|\leq1$, we find
\beq f(t) = \frac{1}{2}(1+\sqrt{1-\gamma^2})
e^{-2\ii (\sqrt{1-\gamma^2}-1)t}.\eeq
The system ground state again exhibits time-crystalline order.
When $\lambda=0$ and $|\gamma|>1$, we obtain
\beq f(t) =
\frac{1}{2} e^{2\ii t}e^{-2\sqrt{\gamma^2-1}~t},\eeq
by choosing $\epsilon_0=-N+2+2\ii\sqrt{\gamma^2-1}$ as the ground state eigen-energy
from the two eigen-values $-N+2\pm2\ii \sqrt{\gamma^2-1}$.
The system ground state now exhibits time functional order phase,
with a decaying $f(t)$ as time evolves.
When $\lambda=\gamma=0$,
\beq f(t) = 1,\eeq
the ground state reduces to time trivial order phase.

The above two non-Hermitian models represent
direct generalizations of the Hermitian system
considered in Refs. \cite{Kozin2019qtcLongRange, Facchi2011GHZ}.
While slightly more complicated,
they remain sufficient simple for compact analytical treatment,
thus helping to reveal interesting and clear physical meanings
of the underline time order.

\subsection{Some remarks about continuous time crystal}
According to the WO no-go theorem \cite{WO2015absence}, $f(t)$ for the ground state or the Gibbs ensemble of
a general many-body Hamiltonian whose interactions are not-too-long ranged
exhibits no temporal dependence, hence belongs to time trivial order according to our classification scheme.
At first sight, this seems to sweep many important models of condensed matter physics
into the same boring class of time trivial order phase.
However, it remains to explore, for instance, many-body systems with
 more than two-body (or $k$-body) interactions, or non-Hermitian systems,
  which might support the existence of CTC.
Inspired by the recent results on CTC \cite{Kozin2019qtcLongRange},
we believe more time crystalline phases will be uncovered
and further understanding will be gained in the future.

As emphasized earlier, continuous time crystal results from spontaneously breaking continuous time translation symmetry.
Due to the genuine time periodicity contained in CTC,
it might be possible to explore and design \textit{new types of clocks}
based on macroscopic many-body systems, as the time
period is directly related to energy spectra, and whose
physical meaning is clearly the same as for atomic clock states.
Furthermore, they are not affected by finite size effect in contrast to periodicity in DTC.

\section{Discussion}
While ground state phases of a quantum many-body system are mostly classified
with its Hamiltonian based on two paradigms: LGW symmetry breaking order parameter
or topological order, this work proposes to study phases from time dimension
using \textit{time order} or more specifically with the proposed symmetry-based time order.
 Compared to the recent progress and understanding gained for topological order \cite{wen2017rmpcollo, wen2019choreographed},
one could try to develop a framework for \textit{entanglement-based time order} instead of the \textit{symmetry-based time order}
we employ here in this study.
Quantum entanglement in a many-body system is responsible for topological order, whose origin lies at the tensor product structure of the quantum many-body Hilbert space
$\mathcal{H}_{\text{tot}}=\otimes_i\mathcal{H}_i$ with $\mathcal{H}_i$ the finite-dimensional Hilbert space for site-$i$.
An \textit{entanglement-based time order} therefore calls for a combined investigation to exploit quantum entanglement
and temporal properties of a quantum many-body system.

Through {\it time order}, one focuses on temporal structure of the evolution operator $e^{-i\hat{H}t}$.
The \textit{symmetry-based time order} therefore unifies LGW paradigm with the concept of {\it time order},
while an
\textit{entanglement-based time order} could amalgamate topological order paradigm (or entanglement beyond that)
with time order. For this to happen,
a more basic definition for {\it time order} will be required,
which will likely expand into further in-depth investigations.

In conclusion, understanding phases of matter constitutes a corner stone of contemporary physics.
Capitalizing on the concept of CTC for many body ground state with perpetual time dependence,
this study argues that information from time domain can be employed to 
classify quantum phase as well, which provides a new perspective
towards the understanding of ground state time dependence,
significantly beyond existing studies on CTC.
We introduce time order, provide its operational definition in terms of two-time auto-correlation 
function of an appropriate symmetry order operator, 
bestow physical meaning to characteristic frequencies and amplitudes of the correlation function,
and present complete classification of
time order phases. 
Time order phase diagrams for a spin-1 BEC system and the quantum Rabi model
are fully worked out. Interesting time order phases 
in non-Hermitian spin models with multi-body interaction are presented.
Besides the time crystalline order which already attracts broad attention from 
its studies in terms of CTC,
other phases we identify, e.g. time quasi-crystalline order and time functional order,
represent exciting new possibilities.

\section{Methods}
The Supplementary Material contains all calculation details.
In Sec. \Rmnum{1} we extend the discussion of time order to finite temperature where concrete examples in spin-1 BEC system are given.
In Sec. \Rmnum{2} we present the numerical method for studying the spin-1 BEC example, while in Sec. \Rmnum{3} we provide the variational result about the polar ground state of a spin-1 BEC.
in Sec. \Rmnum{4} we show details about the ground state calculation in the non-Hermitian quantum many-body models considered.

\section{Acknowledgments}
\begin{acknowledgments}
This work is supported by the National Key R\&D Program of China (Grant
No. 2018YFA0306504), and by the National Natural Science
Foundation of China (NSFC) (Grants No. 11654001 and No. U1930201), and by the Key-Area Research and Development Program of GuangDong Province (Grant No. 2019B030330001).
\end{acknowledgments}

\section{Author contributions}
T.-C.G. proposed and conducted the research, supervised by L.Y.; T.-C.G. and L.Y. discussed the results and wrote the manuscript.

\section{Code availability}
Source code for generating the plots is available from the authors upon request.

\section{Data availability}
The data that support the plots within this paper and other findings of this study are available from the corresponding author upon request.

%

\end{document}


\title{Supplementary Material for ``Quantum Phases of Time Order in Many-Body Ground States''}
\author{Tie-Cheng Guo}
\affiliation{State Key Laboratory of Low Dimensional Quantum Physics,
Department of Physics, Tsinghua University, Beijing 100084, China}
\author{Li You}
\affiliation{State Key Laboratory of Low Dimensional Quantum Physics,
Department of Physics, Tsinghua University, Beijing 100084, China}
\affiliation{Frontier Science Center for Quantum Information, Beijing, China}
\maketitle
\onecolumngrid

This supplementary provides supporting material and related details for 
the presentation of the main text.
It is organized as follows: in Sec.~\ref{sec:sec1} we extend the discussion of time order to finite temperature;
in Sec.~\ref{sec:sec2}, we present calculation details related to the spin-1 atomic Bose-Einstein condensate (BEC) example
considered;
as a more straightforward approach to understand numerical results,
we present a variational approach for treating the polar ground state of a spin-1 BEC in Sec.~\ref{sec:sec3}.
Finally, we give the details about ground state and eigen-energy calculation in the non-Hermitian quantum many-body model with multi-body interaction in Sec.~\ref{sec:sec4}.

\section{Time order at finite temperature}\label{sec:sec1}
At finite temperature $T$, excited states will be populated, which can be
taken into account with the Gibbs ensemble $\hat{\rho}\equiv e^{-\beta\hat{H}}/{Z}$,
where $Z\equiv\text{Tr}~e^{-\beta\hat{H}}$ denotes the partition function
and $\beta\equiv 1/T$ the inverse temperature. We then find
\begin{eqnarray}
	f(t)  	&\rightarrow &  \lim_{V\rightarrow\infty} \text{Tr}\left(e^{i\hat{H}t}\hat{\phi}(0)e^{-i\hat{H}t}\hat{\phi}(0)\hat{\rho}\right)\nonumber
\\
	&=& \lim_{V\rightarrow\infty}\sum_{k=0}^\infty\langle\psi_k|e^{i\hat{H}t}\hat{\phi}(0)e^{-i\hat{H}t}\hat{\phi}(0) \frac{e^{-\beta\hat{H}}}{Z}|\psi_k\rangle\nonumber\\
	&=& \lim_{V\rightarrow\infty}\sum_{k=0}^\infty\frac{1}{Z} e^{i\epsilon_k t-\beta \epsilon_k}\langle v_k| e^{-i\hat{H}t} |v_k\rangle\nonumber\\
	&=& \lim_{V\rightarrow\infty}\sum_{k=0}^\infty\sum_{j=0}^\infty \frac{1}{Z} c_{jk}e^{-\beta \epsilon_k} e^{-i(\epsilon_j-\epsilon_k)t},
\end{eqnarray}
where $|v_k\rangle$ is the eigen-state twisted vector for $|\psi_k\rangle$, and
$c_{jk}$ its associated weight.
Analogously, for the non-Hermitian case, we find
\begin{eqnarray}
	f(t) 
	&=& \lim_{V\rightarrow\infty}\sum_{k=0}^\infty\frac{1}{Z} e^{i\epsilon_k t-\beta \epsilon_k}\langle v_{k}^{(l)}| e^{-i\hat{H}t} |v_{k}^{(r)}\rangle\nonumber\\
	&=& \lim_{V\rightarrow\infty}\sum_{k=0}^\infty\sum_{j=0}^\infty \frac{1}{Z} c_{jk}e^{-\beta \epsilon_k} e^{-i(\epsilon_j-\epsilon_k)t},
\end{eqnarray}
where $|v_{k}^{(l)}\rangle$ and $|v_{k}^{(r)}\rangle$ are the left and right twisted vectors for eigen-state $|\psi_k\rangle$, $c_{jk}$ is the corresponding weight.

It is easily noted that $f(t)$ at finite temperature
contains contributions from all eigen-states of the quantum many-body system $\hat{H}$,
with a temperature dependent weight factor for different energy level, but
$f(t)$ remains to include contributions from different periodic functions. Hence the quantum phase
classification task essentially remains the same (including its possible reference to $F(t)$)
as is shown in the Letter for the ground state.
At finite temperature, due to thermal excitations to ground state,
the temporal behavior will be more complex thus opening up for more interesting possibilities,
e.g., to control {\it time order} phases and
to study crossover or driven phase transitions between different time order phases.

\section{Time order in a spin-1 atomic BEC}\label{sec:sec2}

For typical interaction parameters of a spin-1 BEC (e.g., of ground state $\Rb87$ or $\Na23$ atoms) in a tight trap,
spin domain formation is energetically suppressed when the atom number is not too large as
spin-dependent interaction strength is much weaker than spin-independent interaction \cite{Ho1998, Law1998quantumspins, Ueda2000exact, Ueda2013RMP}.
This facilitates a single-spatial-mode approximation (SMA) by assuming all spin states
share the same spatial wave function $\phi(\bm{r})$, which effectively decouples the spatial degrees of freedom
from the spin and results in the following Hamiltonian \cite{Law1998quantumspins, pu1999spinmixing}
\begin{eqnarray}
  \hat H &= &\frac{c_2}{2N}\left[\left(2\hat N_0 -1\right)\left(\hat N_1+\hat N_{-1}\right)+2\left(\hat a_1^\dag\hat a_{-1}^\dag\hat a_0\hat a_0+\text{h.c.}\right)\right]
  -p \left(\hat N_1 - \hat N_{-1}\right)
	+ \,q\,\left(\hat N_1 + \hat N_{-1}\right)\,,\label{eq:Hamil0}
\end{eqnarray}
for the model many body system,
where $\hat{a}_{m_F}(m_F = 0, \pm 1)$ is the annihilation operator of the ground manifold state $|F = 1, m_F\rangle$
with corresponding number operator $\hat{N}_{m_F} = \hat{a}_{m_F}^\dagger \hat{a}_{m_F}$.
$p$ and $q$ are linear and quadratic Zeeman shifts which could be tuned independently in experiments \cite{luo2017deterministic},
while $c_2$ denotes spin exchange interaction strength.
The total particle number operator $\hat{N} = \hat{N}_1 + \hat{N}_0 + \hat{N}_{-1}$ as well as the
longitudinal magnetization operator $\hat F_z=\hat{N}_1 - \hat{N}_{-1}$ are both conserved.
Thus, linear Zeeman shift can be set to $p=0$ effectively.

As discussed in the main text,
a suitable order parameter for this model system is $\hat{n}_{\rm sum}\equiv \hat{N}_{\rm sum}/{N}$ ($\hat{N}_{\rm sum}={\hat{N}_1+\hat{N}_{-1}}=N-\hat{N}_0$),
which measures the fractional atomic population in the states $|1, 1\rangle$ and $|1, -1\rangle$ and
$N$ assumes the role of system size. Following our formulation and denoting
the system energy eigen-state by $|\psi_i\rangle$ $(i=0,1,2,\cdots)$ with increasing eigen-energy $\epsilon_i$,
the ground state twisted vector becomes $|v\rangle\equiv\hat{{n}}_{\rm sum}|\psi_0\rangle=\sum_{i=0}^\infty a_i|\psi_i\rangle$,
with $a_i=\langle\psi_i|v\rangle$ its expansion coefficient on the eigen-state $|\psi_i\rangle$.
We find
\beq\begin{split}
	f(t) =\lim_{N\rightarrow\infty}\langle\hat{{n}}_{\rm sum}(t)\hat{{n}}_{\rm sum}(0)\rangle 
	= \lim_{N\rightarrow\infty}\sum_{j=0}^\infty b_j e^{-i(\epsilon_j-\epsilon_0)t},
\end{split}\eeq
where $b_j\equiv |a_j|^2$ is the weight of the ground state twisted vector, $b\equiv\sum_{j=0}^\infty b_j$ the total weight, and
\begin{eqnarray}
	F(t) = \lim_{N\rightarrow\infty}
	\left\langle\hat{N}_{\rm sum}(t)\hat{N}_{\rm sum}(0)\right\rangle 
	= \lim_{N\rightarrow\infty}\sum_{j=0}^\infty B_j e^{-i(\epsilon_j-\epsilon_0)t},
\end{eqnarray}
where $A_i=N\langle\psi_i|v\rangle$, $B_j\equiv |A_j|^2$ is the weight of the enlarged ground state twisted vector,
and $B\equiv\sum_{j=0}^\infty B_j$ the total weight.

Our study below is for the zero magnetization $F_z=0$ subspace
and employs exact diagonalization (ED) to calculate eigen-states as well as eigen-energies.
The overall time order phase diagram for spin-1 BEC is shown in the main Letter.
For ferromagnetic interaction $c_2<0$, the critical quadratic Zeeman energy $q/|c_2|=2$
splits the whole region into time trivial order (TT) phase for smaller $q$ that observes TTS,
and the generalized time crystalline (gTC) order phase for $q/|c_2|>2$ where TTS is spontaneously broken.
The latter (gTC phase) is found to coincide with the ground state polar phase.
The available computation resource limits the calculation to a finite system size,
which prevents us from mapping out the exact details in the immediate neighborhood of $q=2$,
where further elaboration is need for its time order properties.
On the other hand, for antiferromagnetic interactions, we find $q=0$ separates TT phase from and gTC order.

\begin{figure}[htbp]
\centering
\includegraphics[scale=1]{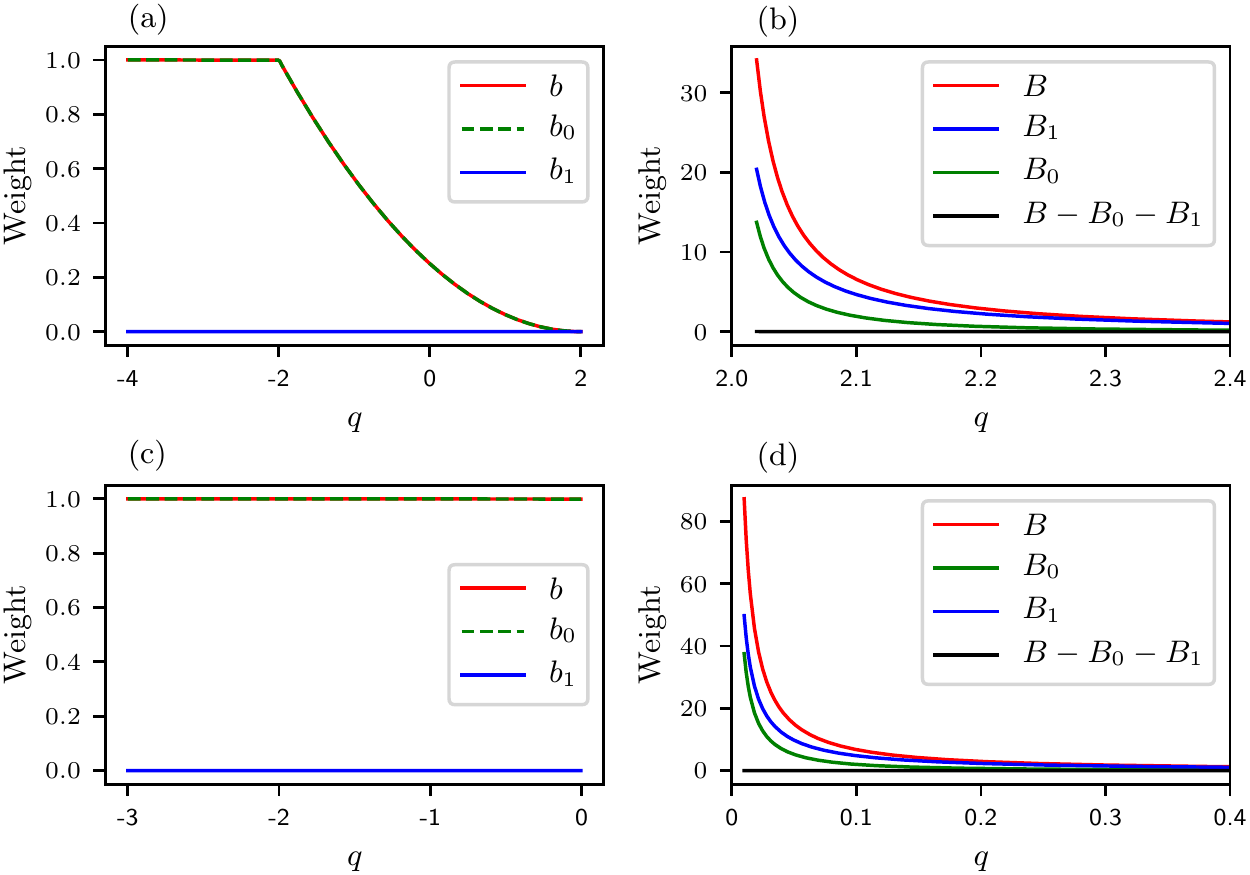}
	\caption{Weights of ground state twisted vector in the ground and low-lying excited states
as functions of $q$ at system size $N=10000$.
	The upper panel is for ferromagnetic interaction, where
	weights $b_i$ for $q<2$ are shown in (a) while weights $B_i$ for $q>2$ are shown in (b).
	The lower panel is for antiferromagnetic interaction, where
	weights $b_i$ for $q<0$ are shown in (c) while weights $B_i$ for $q>0$ are shown in (d).
}
\label{fig:c_q}
\end{figure}

In Figure \ref{fig:c_q}, the weights for the ground state as well as for the low-lying excited states
are shown as functions of $q$ for a typical system size of $N=10000$.
Only the ground state weight $b_{0}$ is nonvanishing in the $q<2$ ($q<0$) region
for ferromagnetic (antiferromagnetic) interactions,
but the total weight $b$ is zero in the $q>2$ ($q>0$) region for ferromagnetic (antiferromagnetic) interaction,
which prompts us to examine further the enlarged weights $B_i$ corresponding to the bulk order parameter.
For ground and the first excited states, the volume enlarged weights $B_{0,1}$ are found to be nonvanishing,
although both decrease as $q$ increases and grow with $N$
as $q$ approaches $q=2$ ($q=0$) for ferromagnetic (antiferromagnetic) interaction.
However, as mentioned above, limited to a system size of $N=10000$ by computation resource in the ED calculation, we cannot exactly map out the behavior near $q=2$ ($q=0$) for ferromagnetic (antiferromagnetic) interaction.
This consequently leaves empty for $q$ in region $[2.0, 2.02]$ ($[0, 0.01]$) for ferromagnetic (antiferromagnetic) interaction.

\begin{figure}[htbp]
\centering
\includegraphics[scale=1]{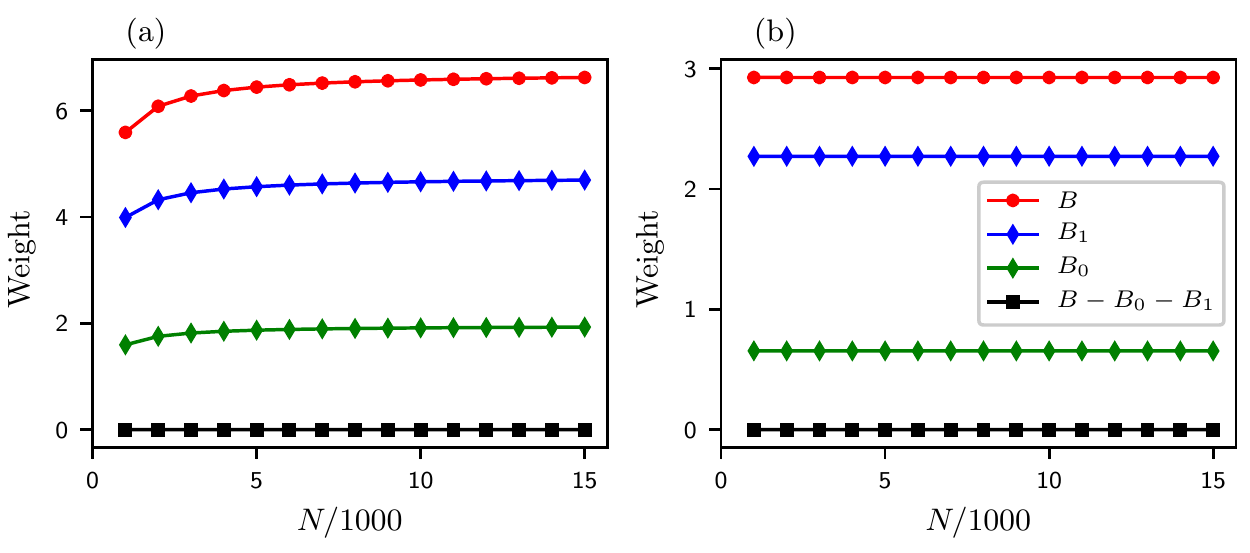}
\caption{Weights of ground state twisted vector in the ground and low-lying excited states
as functions of system size $N$ at $q=2.1$ for ferromagnetic interaction (a)
and at $q=0.2$ for antiferromagnetic interaction (b).
	}
\label{fig:c_N}
\end{figure}

The dependence on system size $N$ is clearly revealed by Fig. \ref{fig:c_N},
with the enlarged weights in the gTC regime attain fixed values as system approaches thermodynamic limit ($N\to\infty$).
In regions away from $q=2$ ($q=0$) for ferromagnetic (antiferromagnetic) interaction, 
ED numerics can always approach thermodynamic limit except for the
immediate neighbourhood near $q=2$ ($q=0$), where we infer with confidence the tendencies to divergence
of the weights $B_{0,1}$ as $q$ approaches $q=2$ ($q=0$).

\begin{figure}[htbp]
\centering
\includegraphics[scale=1]{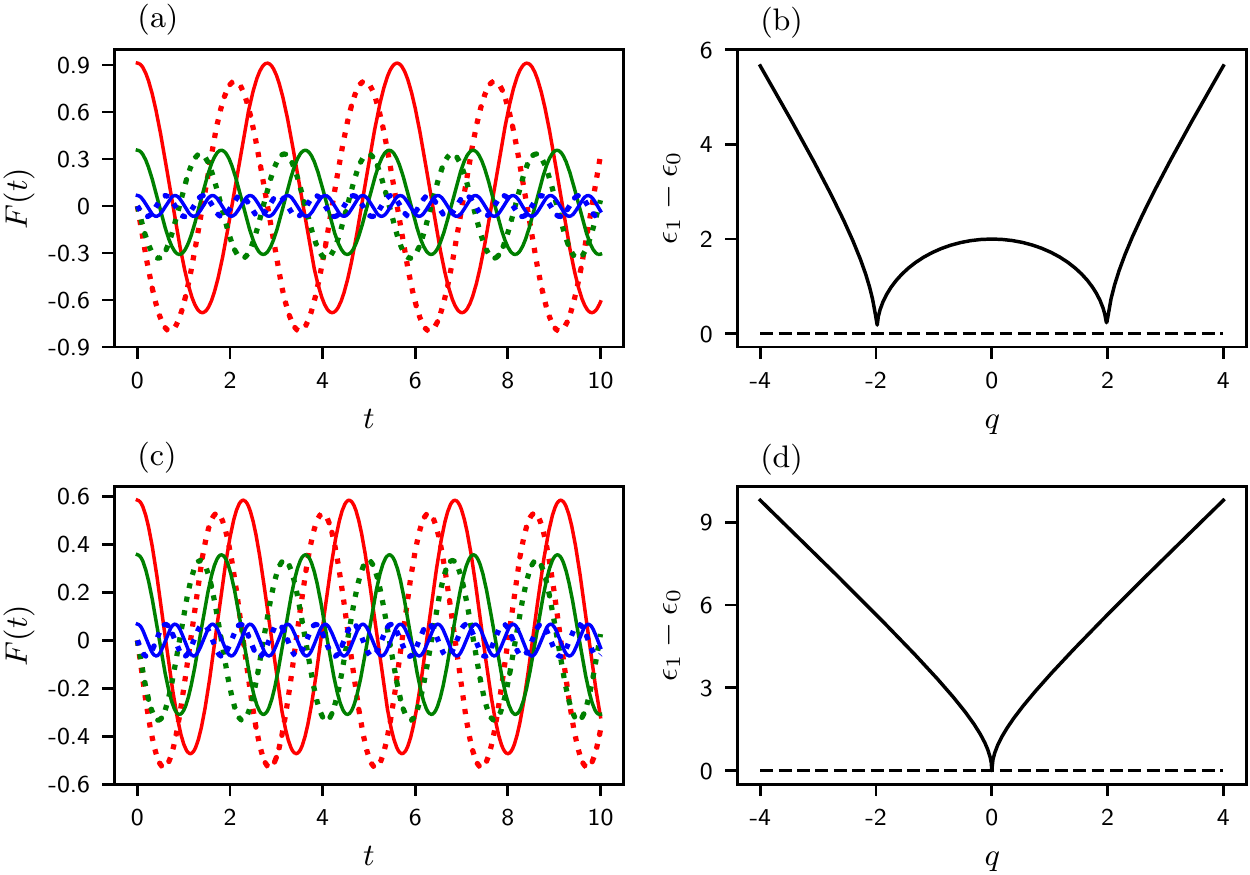}
	\caption{$F(t)$ for different $q$ as a function of time $t$. The solid and dotted lines correspond to ${\rm Re}(F)$ and ${\rm Im}(F)$ respectively. The red, green, and blue lines {correspond} to $q=2.5$, $q=3$, and $q=5$ respectively for ferromagnetic interaction (a). The red, green, and blue lines correspond to $q=0.7$, $q=1$, and $q=3$ respectively for antiferromagnetic interaction (c). The energy gap between ground and the first excited state $\epsilon_1-\epsilon_0$ as a function of $q$ for ferromagnetic (b) and antiferromagnetic interactions (d), at system size $N=5000$.}
\label{fig:Ft}
\end{figure}

The time evolution of two-time auto-correlation function $F(t)$ is plotted in Fig. \ref{fig:Ft}
(a) for ferromagnetic and (c) for antiferromagnetic interactions,
while Figs. \ref{fig:Ft} (b) and (d) display energy gaps between ground and the first excited states
as a function of $q$ for ferromagetic and antiferromagnetic interactions respectively
at a system size of $N=5000$.
The behavior of $F(t)$ is quantitatively consistent with that of the weights $B_i(q)$ $(i=0, 1)$
shown in Fig. \ref{fig:c_q} and the energy gap $\epsilon_1-\epsilon_0$ shown in Figs. \ref{fig:Ft} (b) and (d).

\begin{figure}[htbp]
\centering
	\includegraphics[scale=1]{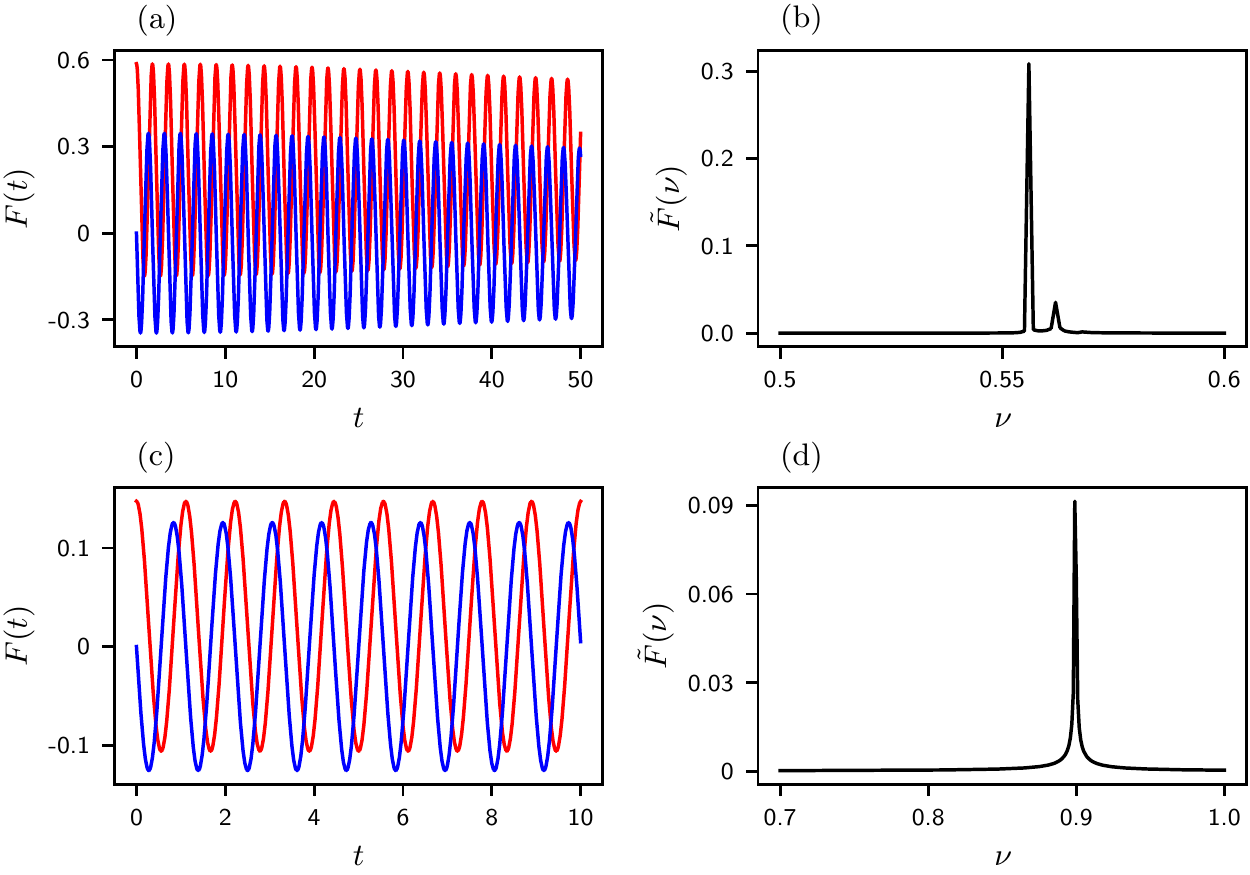}   
	\caption{$F(t)$ as a function of time $t$ at $q=3$ for ferromagnetic interaction (a) and $q=2$ for antiferromagnetic interaction (c). The red and blue solid lines respectively correspond to ${\rm Re}(F)$ and ${\rm Im}(F)$. The Fourier transform spectrum $\tilde{F}(\nu)$ of ${\rm Re}(F)$ or ${\rm Im}(F)$ with $\nu=1/T$ the frequency, $T$ the period, for ferromagnetic (b) and antiferromagnetic (d) interactions,
at temperature $\beta=1$ and system size $N=500$.}
\label{fig:thermal}
\end{figure}

At finite temperature, excited states come into play by also contributing to the correlation function.
We find the gTC order hosted in the polar phase persists for both ferromagnetic
and antiferromagnetic interactions.
The corresponding time evolution and Fourier transform of $F(t)$ are shown in Fig. \ref{fig:thermal},
calculated for $N=500$ at a temperature of $\beta\equiv1/T=1$.
The Fourier transform is performed for ${\rm Re}(F)$ over $t=[0,1000]$ with the zero frequency (DC) component subtracted or for ${\rm Im}(F)$.
The upper (lower) panel corresponds to ferromagnetic (antiferromagnetic) interaction at $q=3$ ($q=2$).
For ferromagnetic interaction, two distinct frequency components are clearly identified for $q=3$,
associated with the two different energy level gaps.
Thus, the gTC phase remains at finite temperature.
Moreover, we also find a generalized time quasi-crystalline order phase assuming
the two frequencies are incommensurate, by fine tuning their corresponding energy gaps such that
the relation $\Delta_1/\Delta_2=m_1/m_2$ with $m_1$ and $m_2$ being co-primes is not satisfied.
The gTC phase at finite temperature here is robust which is in contrast
to the melting behavior of Continuous Time Crystal (CTC) shown in Ref. \cite{Kozin2019qtcLongRange}.

Finally, we hope to address the critical question about how could this time order, sort of a perpetual time dependence, can be observed.
 We note the bulk two-time auto-correlation function introduced
$F(t) = \lim_{N\rightarrow\infty}
	\left\langle\hat{N}_{\rm sum}(t)\hat{N}_{\rm sum}(0)\right\rangle$
denotes nothing but the ground state (averaged) conditional outcome of measuring $N_{\rm sum}(t)$ at $t$
after starting with $N_{\rm sum}(0)$ initially. The dynamics of $F(t)$ follows that of
$N_{\rm sum}(t)$ as in quantum regression theorem. Given the system is well controlled, highly reproducible,
one can simply detect $F(t)$ by measuring $N_{\rm sum}(t)$, although for each measurement
at an instant $t$, a condensate is destroyed, and a follow up one will have to be prepared
as closely as possible in every respects (through selection and post-selection) 
and be measured at a different $t'>t$.
Thus, a plausible way to detecting the ground state time dependence will require reconstructing
the time dependence of $F(t)/N_{\rm sum}(0)$. As along as the oscillation amplitude is
more than a few percent, it will be easily observable with not too much difficulty,
although such a reconstruction will still be difficult as $N_{\rm sum}(0)$ can be
rather small compared to $N_0\sim N$ in the polar state.
Alternatively, one can perhaps start from a twin-Fock state, i.e., by preparing an initial
state with $N_{\rm sum}(0)\sim N$.

\begin{figure}[htbp]
\centering
\includegraphics[scale=1]{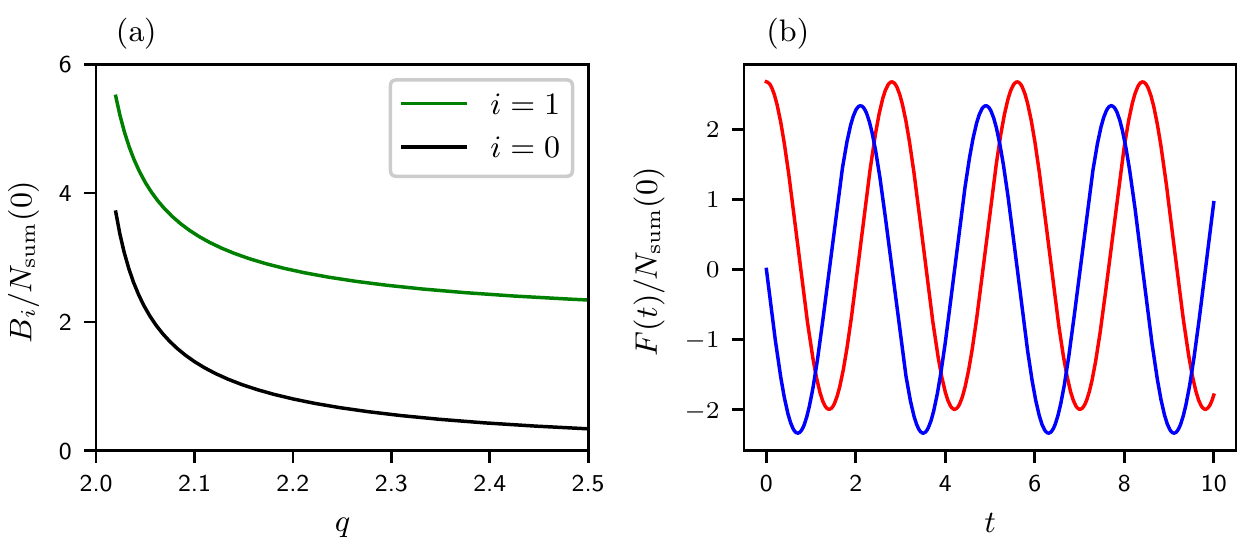}
	\caption{(a) $B_i/N_{\rm sum}(0)$ as a function of $q$ for ferromagnetic interaction. (b) $F(t)/N_{\rm sum}(0)$ as a function of time $t$ at $q=2.5$ for ferromagnetic interaction. The red and blue solid lines correspond to the real and imaginary part of $F(t)/N_{\rm sum}(0)$ respectively.}
\label{fig:contract}
\end{figure}

In Figure \ref{fig:contract}(a) we show the behavior of oscillation amplitude for $F(t)/N_{\rm sum}(0)$.
The time dependence of $F(t)/N_{\rm sum}(0)$ at $q=2.5$ for ferromagnetic interaction is shown in Fig. \ref{fig:contract}(b).

\section{A variational polar state for ferromagnetic spin-1 BEC}\label{sec:sec3}
One might naively expect that nothing particularly interesting could happen in the polar phase of a ferromagnetic spin-1 BEC,
 where essentially all atoms reside in the single particle state $|1, 0\rangle$.
Nevertheless, due to the competition between spin exchange interaction $c_2$ and quadratic Zeeman shift $q$,
the ground state of our system differs from $|N_1=0, N_0=N, N_{-1}=0\rangle$, 
which can be affirmed based on a simple
variational analytical calculation given in this section.

We use the number state basis $|N_1, N_0, N_{-1}\rangle\equiv|[N], M, k\rangle$, where $N_{m_F}$ denotes occupation number of the $m_F$ magnetic state, $M\equiv N_{1}-N_{-1}$, and $k\equiv N_{-1}$. We take the following ground state variational ansatz
$|\psi_0\rangle = \frac{1}{\sqrt{1+|a|^2}}\left(|0, N, 0\rangle + a|1, N-2, 1\rangle\right)$ for the polar state of ferromagnetic spin-1 BEC,
where $a=r e^{i\phi}$ is a (complex) variational parameter with $r$ and $\phi$ real.
From Eq. \eqref{eq:Hamil0} and (assumed) $p=0$, the ground state energy follows from
\begin{eqnarray}
E & = &\frac{\langle\psi_0|H|\psi_0\rangle}{\langle\psi_0|\psi_0\rangle}\nonumber\\
	&= &\frac{1}{1+a^*a}
	\Big[
		\big(\frac{c_2(2N-5)}{N} + 2q\big)a^*a
		+ c_2\sqrt{\frac{N-1}{N}}(a^*+a)
		\Big]\nonumber\\
&=& \frac{1}{1+r^2}
	\Big[
		\big(\frac{c_2(2N-5)}{N} + 2q\big)r^2
		+ 2 c_2\sqrt{\frac{N-1}{N}}\, r\cos(\phi)
		\Big].
\end{eqnarray}
We see the extreme value (the minimum) of $E$ is reached when $\cos(\phi)=\pm 1$, i.e., for a real variational parameter $a$,
which will be assumed from now on.  This gives
\beq
E= \frac{x_1 a^2 + x_2 a}{1+a^2},
\eeq
with
$x_1=\frac{c_2(2N-5)}{N} + 2q$ and $x_2=2c_2\sqrt{\frac{N-1}{N}}$.
The derivative of the energy function $E(a)$ is
\beq E^\prime(a) = \frac{-x_2 a^2 + 2 x_1 a + x_2}{(1+a^2)^2},\eeq
which determines the locations for the extreme values
\begin{eqnarray}
a_\pm &=& \frac{1}{2c_2\sqrt{N(N-1)}}\Big[c_2(2N-5) + 2Nq
\pm N\sqrt{\frac{c_2^2(8N^2-24N+25)}{N^2} + \frac{4c_2q(2N-5)}{N} + 4q^2}\Big],
\end{eqnarray}
and the corresponding extreme values are
\begin{eqnarray}
E_\pm = c_2 + q\pm \frac{1}{2}\sqrt{4q^2+8c_2q+8c_2^2 +\frac{-24c_2^2-10c_2q}{N} + \frac{25c_2^2}{N^2}} - \frac{5 c_2}{2 N}.
\end{eqnarray}

In the thermodynamic limit $N\rightarrow\infty$, they reduce respectively to
 $a_\pm = 1+\frac{q}{c_2}\pm\frac{\sqrt{2c_2^2+2c_2q+q^2}}{c_2}$ and
 $E_\pm = c_2 + q \pm\sqrt{2c_2^2 + 2c_2q + q^2}$.
The left and right asymptotic value for the energy function $E(a)$ is therefore
\beq E(a) = c_2(2-\frac{5}{N})+2q,\ \ (\text{when}\, a\rightarrow\pm\infty).\eeq

For ferromagnetic interaction $(c_2<0)$, $E_-$ assumes the minimum, which corresponds to
the ground state $|\psi_0\rangle = \frac{1}{\sqrt{1+a_-^2}} \left(|0, N, 0\rangle + a_- |1, N-2, 1\rangle \right)$ with $N_{\rm sum}=2 a_-^2/(1+ a_-^2)$, and $a_-=1+\frac{q}{c_2}-\frac{\sqrt{2c_2^2+2c_2q+q^2}}{c_2}$
in the thermodynamic limit $N\rightarrow \infty$.

Despite of the vanishing order parameter $n_{\rm sum}$ in the polar phase (here the gTC order phase from the time order perspective),
the enlarged quantity $N_{\rm sum}$ retains a finite value.
Hence, the physics we present here clearly belongs to the realm of quantum effects,
 beyond the reach of mean-field theory.

\section{The non-Hermitian spin model with multi-body interaction}\label{sec:sec4}
The non-Hermitian quantum many body model Hamiltonian is
\beq \hat{H}= \hat{H}_0 + (\lambda+\gamma\ii )\hat{H}_1,\eeq
with
\beq\begin{split}
	\hat{H}_0 &= -\sum_{j=1}^N\sigma_j^z\sigma_{j+1}^z,\\
	\hat{H}_1 &= \sigma_1^x\sigma_2^x\cdots\sigma_{[N/2]}^x
-\sigma_{[N/2]+1}^x\cdots\sigma_N^x,
\end{split}\eeq
where $[\cdot]$ denotes the integral part, $\bm\sigma_{N+1}\equiv\bm\sigma_1$, $\lambda$ and $\gamma$ are spin-string interaction strength and dissipation strength respectively. $\lambda$ and $\gamma$ are both real numbers. $\ii$ is the imaginary unit. $\sigma^{x,y,z}$ are Pauli operators. $N$ is the qubit number of the system.
The Hamiltonian has the $[(N+1)/2]-$ body interaction term and supports the GHZ state $|G_+\rangle$ as a non-degenerate excited state. 

Firstly, denote the Greenberger-Horner-Zeilinger (GHZ) states as
\beq |G_\pm\rangle = \frac{1}{2}(|0\rangle^{\otimes N} \pm |1\rangle^{\otimes N}).\eeq
Denote
\beq |\tilde{G}_{-,\mathcal{I}}\rangle = \frac{1}{\sqrt{2}}\left(
|0\rangle_1\cdots|0\rangle_{[N/2]}|1\rangle_{[N/2]+1}\cdots|1\rangle_N
-|1\rangle_1\cdots|1\rangle_{[N/2]}|0\rangle_{[N/2]+1}\cdots|0\rangle_N
\right),
\eeq
where $\mathcal{I}=([N/2]+1, [N/2]+2, \cdots, N)$ is a multi-index.

We immediately know that $|G_\pm\rangle$ are the degenerate ground state of the ferromagnetic Ising Hamiltonian $\hat{H}_0$ with eigen-energy $E_{(0)}=-N$, $|\tilde{G}_{-,\mathcal{I}}\rangle$ is the excited state of $\hat{H}_0$ with eigen-energy $E_{(1)} = -N+4$.

The action of $\hat{H_1}$ on $|G_-\rangle$ ($|\tilde{G}_{-,\mathcal{I}}\rangle$) gives $|\tilde{G}_{-,\mathcal{I}}\rangle$ ($|G_-\rangle$) with a multiplicative factor $-2$.
\beq\begin{split}
\hat{H}_1|G_-\rangle&=-2|\tilde{G}_{-,\mathcal{I}}\rangle,\\
\hat{H}_1|\tilde{G}_{-,\mathcal{I}}\rangle & = -2|G_-\rangle.
\end{split}\eeq
Then we know the two eigen-states of $\hat{H}$ is a superpositon of $|G_-\rangle$ and $|\tilde{G}_{-,\mathcal{I}}\rangle$ and can be written as 
\beq |\Psi\rangle= \alpha_1|G_-\rangle+\alpha_2|\tilde{G}_{-,\mathcal{I}}\rangle,\eeq
where $\alpha_{1,2}$ are the undetermined coefficients.
Substitute into the Schr\"odinger equation $\hat{H}\Psi\rangle = \epsilon|\Psi\rangle$, we get
\begin{align}
	\epsilon^2-\left(E_{(0)}+E_{(1)}\right)\epsilon+\left(E_{(0)}E_{(1)}-4(\lambda+\gamma\ii)^2\right) & = 0,\\
	2(\lambda+\gamma\ii)\alpha_2 & = (E_{(0)}-\epsilon)\alpha_1.
\end{align}
The eigen-energy $\epsilon^{(\pm)}=\frac{1}{2}[E_{(0)}+E_{(1)}\pm\sqrt{(E_{(1)}-E_{(0)})^2+16(\lambda+\gamma\ii)^2}]$. Choose $\alpha_1=1$, we have $\alpha_2 = \frac{E_{(0)}-\epsilon}{2(\lambda+\gamma \ii)}$. Imposing the normalization condition, we have
\beq |\Psi^{(\pm)}\rangle= \frac{\alpha_1}{\sqrt{|\alpha_1|^2+|\alpha_2|^2}}|G_-\rangle+\frac{\alpha_2}{\sqrt{|\alpha_1|^2+|\alpha_2|^2}}|\tilde{G}_{-,\mathcal{I}}\rangle.\eeq

If $\gamma=0$, the Hamiltonian is Hermitian and we have the ground state $|\Psi_0\rangle\equiv|\Psi^{(-)}\rangle$ with energy $\epsilon_0 \equiv\epsilon^{(-)}=-N-2(\sqrt{1+\lambda^2}-1)$.
See more details about the Hermitian version of the system in Ref. \cite{Facchi2011GHZ}.
Here, we choose the eigen-state from $\{|\Psi^{(+)}\rangle, |\Psi^{(-)}\rangle\}$ as the ground state $|\Psi_0\rangle$ of our generalized non-Hermitian system, for it deforms into the ground state of the Hermitian case when $\gamma$ approaches zero.
If $\epsilon$ is real, then the ground state energy $\epsilon_0$ corresponds to the smaller one from $\epsilon^{(\pm)}$.
However, the ground state energy $\epsilon_0$ corresponds to the one with the larger imaginary part when $\epsilon$ is a complex number.
The ground state $|\Psi_0\rangle$ are obtained straightforwardly.

For the GHZ state $|G_+\rangle$, we can know it's a non-degenerate excited state with energy $\epsilon_+ = -N$, for
\beq \hat{H}_1 |G_+\rangle = 0.\eeq

%